# Competitive co-adsorption of $CO_2$ with $H_2O$, $NH_3$, $SO_2$, $NO$, $NO_2$, $N_2$, $O_2$, and $CH_4$ in M-MOF-74 (M= Mg, Co, Ni): the role of hydrogen bonding

Kui Tan,[†] Sebastian Zuluaga,[‡] Qihan Gong,[§] Yuzhi Gao,[†] Nour Nijem,[†±] Jing Li,[§] Timo Thonhauser,[‡] Yves J Chabal[†]*

[†]Department of Materials Science & Engineering, University of Texas at Dallas, Richardson, Texas 75080

[‡]Department of Physics, Wake Forest University, Winston-Salem, North Carolina 27109

[§]Department of Chemistry and Chemical Biology, Rutgers University, Piscataway, New Jersey 08854

[±] Department of Chemistry, University of California Berkeley, Berkeley, California, 94720

Supporting Information

**ABSTRACT:** The importance of co-adsorption for applications of porous materials in gas separation has motivated fundamental studies, which have initially focused on the comparison of the binding energies of different gas molecules in the pores (i.e. energetics) and their overall transport. By examining the competitive co-adsorption of several small molecules in M-MOF-74 (M= Mg, Co, Ni) with *in-situ* infrared spectroscopy and *ab initio* simulations, we find that the binding energy at the most favorable (metal) site is not a sufficient indicator for prediction of molecular adsorption and stability in MOFs. Instead, the occupation of the open metal sites is governed by kinetics, whereby the interaction of the guest molecules with the MOF organic linkers controls the reaction barrier for molecular exchange. Specifically, the displacement of $CO_2$ adsorbed at the metal center by other molecules such as $H_2O$, $NH_3$, $SO_2$, $NO$, $NO_2$, $N_2$, $O_2$, and $CH_4$ is mainly observed for $H_2O$ and $NH_3$, even though $SO_2$, $NO$, and $NO_2$, have higher binding energies (~70-90 kJ/mol) to metal sites than that of $CO_2$ (38 to 48 kJ/mol) and slightly higher than water (~60-80 kJ/mol). DFT simulations evaluate the barriers for $H_2O \rightarrow CO_2$ and $SO_2 \rightarrow CO_2$ exchange to be ~ 13 and 20 kJ/mol, respectively, explaining the slow exchange of $CO_2$ by $SO_2$, compared to water. Furthermore, the calculations reveal that the kinetic barrier for this exchange is determined by the specifics of the interaction of the second guest molecule (e.g., $H_2O$ or $SO_2$) with the MOF ligands. Hydrogen bonding of $H_2O$ molecules with the nearby oxygen of the organic linker is found to facilitate the positioning of the $H_2O$ oxygen atom towards the metal center, thus reducing the exchange barrier. In contrast, $SO_2$ molecules interact with the distant benzene site, away from the metal center, hindering the exchange process. Similar considerations apply to the other molecules, accounting for much easier $CO_2$ exchange for $NH_3$ than for $NO$, $NO_2$, $CH_4$, $O_2$, and $N_2$ molecules. In this work, critical parameters such as kinetic barrier and exchange pathway are first unveiled and provide insight into the mechanism of competitive co-adsorption, underscoring the need of combined studies, using spectroscopic methods and *ab initio* simulations to uncover the atomistic interactions of small molecules in MOFs that directly influence co-adsorption.

## 1. Introduction

Co-adsorption and gas separation in mixtures is important for many energy and environmental applications such as $CO_2$ capture, $CH_4/H_2$ purification, desulfurization (DeSO$_x$) and denitrification (DeNO$_x$).[1-5] The mechanisms are complex because they involve equilibrium adsorption and mixture kinetics.[6] For small molecules with similar sizes, equilibrium adsorption based on binding energies is considered to be one of the major factors to determine the selectivity. For example, selective adsorption of one species over another is attributed to differences in the individual guest molecule affinity to the adsorbent surface.[1,3,6] Some parameters such as adsorption uptake, enthalpy and selectivity can be estimated from the adsorption isotherms (i.e., assuming equilibrium), providing a simple way to evaluate the preferential adsorption of mixture gases in a particular material.[3,6] However, adsorption isotherms are often limited to single component analysis and do not take into account the competition of different molecules for specific adsorption sites.[3]

Competitive adsorption is a common phenomenon for gas adsorption. However, mechanistic studies of competitive co-adsorption have mostly involved flat surfaces of *nonporous* materials such as metal, metal oxides and graphite surfaces.[7-14]

Atomic-scale characterization of various gas molecules in porous materials, such as metal organic frameworks (MOFs), is lacking due in part to experimental difficulties. Such studies are critical to the development of gas separation processes.[3,6,15,16] It is likely, for instance, that kinetic factors are important and need to be studied

The only way to experimentally measure the competitive co-adsorption and investigate the kinetics is to monitor the actual occupation of specific adsorption sites for different mixtures as a function of parameters such as time, temperature, and pressure. In this respect, the interpretation of isotherm measurements is particularly challenging for gas mixtures, since the uptake of each adsorbed species can be only inferred indirectly by determining the composition of gas phase evolved. The method is also limited by sample size (i.e., amount of adsorption for the target materials).[15] Alternatively, breakthrough experiments offer a relatively straightforward way to measure muticomponent systems. In the typical set-up for breakthrough measurement, the gas mixture flows through a packed sample and the concentration at the outlet is measured as a function of time.[3,15] The dynamic gas adsorption capacity can be determined by relating the time it takes to breakthrough to relevant molecular adsorption parameters.

However, such measurements do not provide molecular-scale knowledge of competitive co-adsorption.[15,17-20] Infrared spectroscopy, on the other hand, can provide information on the local binding configuration of gas molecules, including mixtures. For instance, we have determined the local adsorption sites of individual molecules ($H_2$, $CO_2$, $SO_2$, and $H_2O$) in M-MOF-74 (M= Zn, Mg, Ni, Co).[21-25] The integrated areas of absorption bands can be used to determine the concentration of each species adsorbed in MOF sample, i.e. the specific IR bands are correlated with the amount (loading) of adsorbed molecules if the dipole moment of molecules does not vary as the molecules are adsorbed, as is the case of all the molecules considered here.[22,26]

In the present work, we use $CO_2$ as a probe molecule to study competitive co-adsorption of $CO_2$ with a variety of small molecules ($H_2O$, $NH_3$, $SO_2$, $NO$, $NO_2$, $N_2$, $O_2$, and $CH_4$) in M-MOF-74 (M= Mg, Co, Ni). This MOF compound is particularly interesting because it is an excellent material for $CO_2$ capture, with the highest $CO_2$ adsorption capacity among MOFs materials.[4,16,27-29] Techniques such as neutron powder diffraction and X-ray diffraction have shown that $CO_2$ molecules adsorb primarily onto the open metal sites through coordination of one of its oxygen to the metal ions.[28,30-34] The adsorption enthalpy determined by isotherm measurements ranges from 38 to 48 kJ/mol for M-MOF-74 (M= Zn, Mg, Ni, Co) materials, depending on the metal center.[4] While experimental and theoretical studies of adsorption of pure $CO_2$ in M-MOF-74 (M= Zn, Mg, Ni, Co) have been well established, co-adsorption with relevant gases (i.e., flue gas contaminants such as $H_2O$, $SO_2$, $NO$, $NO_2$) remains largely unexplored and only limited to a few theoretical studies.[20,35-37] For $CO_2$ capture by MOF-74 materials, exchange of $CO_2$ through competitive adsorption would in principle be possible with other molecules such as $H_2O$, $SO_2$, $NO$, $NO_2$, since all the molecules preferentially adsorb onto the open metal sites with higher binding energies than $CO_2$.[35,36,38-42] However, the confinement and the specific chemical environment within the nanopores of the MOFs can modify the molecular interactions by introducing unexpected kinetic barriers. It is therefore critical to determine experimentally whether $CO_2$ pre-adsorbed at metal sites remains stable in the presence of, or is replaced by these other molecules. To that end, we have combined *in-situ* infrared spectroscopy with DFT calculations to study the complex co-adsorption behavior of $CO_2$ with $H_2O$, $NH_3$, $SO_2$, $NO$, $NO_2$, $N_2$, $O_2$, and $CH_4$ in M-MOF-74 (M= Mg, Co, Ni).

Briefly, we find that the ability for pre-adsorbed $CO_2$ to withstand the attack (i.e., displacement) by another molecule ($H_2O$, $NH_3$, $SO_2$, $NO$, $NO_2$ or $N_2$, $O_2$, $CH_4$) depends on the chemical nature of these other molecules and their interaction with the MOF's linkers. From the groups of guest molecules, only $H_2O$ and $NH_3$ are able to displace $CO_2$ readily. Surprisingly, molecules with higher binding energies at the metal sites, such as $SO_2$, $NO$, and $NO_2$, cannot efficiently remove $CO_2$. Of course, other molecules that weakly (< 30 kJ/mol) bind to the metal centers, $N_2$, $O_2$, and $CH_4$, cannot replace the pre-adsorbed $CO_2$.

To understand these experimental results, DFT calculations have been performed to elucidate the reaction pathways and evaluate the kinetic barriers. For the $H_2O \rightarrow CO_2$ system, it was found that $H_2O$ molecules (and even $NH_3$ molecules) are initially adsorbed on the nearby oxygen sites in close proximity to the pre-adsorbed $CO_2$ molecule by hydrogen bonding, which facilitates the displacement of the $CO_2$ molecule. In contrast, for the $SO_2 \rightarrow CO_2$ system, the $SO_2$ molecule either remains in the channel or binds to the organic linker away from the metal center. As a result, the $SO_2 \rightarrow CO_2$ exchange process is energetically more costly (20 kJ/mol) than the $H_2O \rightarrow CO_2$ (13 kJ/mol). The larger kinetic barrier is responsible for the relatively slow exchange process of $CO_2$ by $SO_2$.

## 2. Experimental and Computational details
### Synthesis of MOF samples:
The synthesis of M-MOF-74 [M= Mg, Ni, Co] samples and their physical characterization (X-ray diffraction and adsorption isotherm) are described in Section 1 of the Supporting Information. These recipes and measurements follow the standard procedures reported in literatures.

### Infrared spectroscopy:
A powder of M-MOF-74 (M= Mg, Co, Ni) sample (~2 mg) was gently pressed onto a KBr pellet (~1 cm diameter, 1-2 mm thick) surface with a pressure of ~6000 to 6500 psi (or a tungsten mesh for $NO_2$ gas studies since $NO_2$ reacts with KBr) and placed into a high-pressure high-temperature cell (product number P/N 5850c, Specac Ltd, UK) at the focal point of the sample compartment of an infrared spectrometer (Nicolet 6700, Thermo Scientific, US). The retention of the MOFs crystal structure after pressing the powder onto the KBr pellets was confirmed by Raman spectra of the samples, as shown in the Figure S3. The samples were activated in vacuum (base pressure < 20 mTorr) at 180 °C for at least 3 h. When IR measurements showed that $H_2O$ pre-adsorbed during sample preparation was fully removed, the sample was cooled back to lower temperatures (24 °C, 50 °C and 75 °C depending on the experiment) for measurements.

*Pure $CO_2$ adsorption into M-MOF-74 (M= Mg, Co, Ni):* Since the IR absorption of $CO_2$ gas at pressures above 20 Torr is prohibitively high (no signal on the detector), it is not possible to distinguish the adsorbed $CO_2$ from gas phase $CO_2$ (see Figure S3). To study the $CO_2$ adsorption and desorption rate, the activated sample was first exposed to 80 Torr $CO_2$ (close to the partial pressure of $CO_2$ in flue gas) and the cell was evacuated by pumping, taking ~ 20 min to reach 10-30 mTorr. The spectra were recorded as a function of time during the desorption process.

*Co-adsorption of $CO_2$ and $H_2O$, $NH_3$, $SO_2$, $NO$, $NO_2$, $N_2$, $O_2$, and $CH_4$:* After $CO_2$ loading and subsequent pumping down to below 200 mTorr, pumping was stopped and other gases (e.g., $H_2O$, $NH_3$, $SO_2$, $NO$, $NO_2$, $O_2$, $N_2$ and $CH_4$) were introduced at pressures specified in the following section. The spectra were recorded as a function of time after introducing gases to the cell to monitor the co-adsorption of loaded gases with pre-adsorbed $CO_2$.

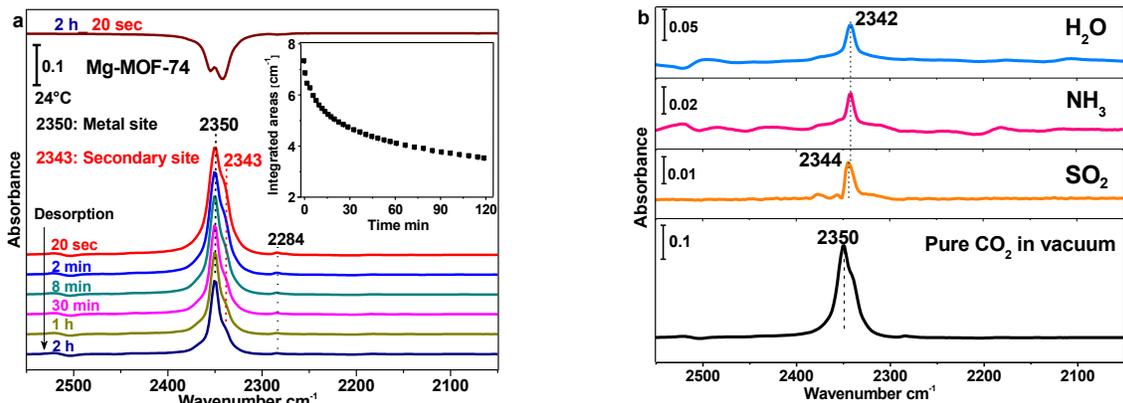

**Figure 1.** (a) IR absorption spectra of $CO_2$ desorption from Mg-MOF-74 as a function of time. Top spectra (dark yellow) in (a) shows the differential spectrum after evacuation for 2 h referenced to the spectra recorded in the very beginning (20 sec). Inset shows the integrated areas for the $CO_2$ $v_3$ band change within 2 h. The gas phase (< 200 mTorr) in the very beginning (< 2 min) was subtracted. (b) IR spectra of $CO_2$ adsorbed into the Mg-MOF-74 pre-exposed to $H_2O$, $NH_3$, and $SO_2$, compared to the $CO_2$ adsorption into activated MOFs.

**DFT calculations:**

The *ab initio* calculations in this work were performed using density functional theory with a plane-wave basis, as implemented in the VA5.3.3 code.[43-45] To take into account the important van der Waals interactions the van der Waals density functional vdW-DF[46-48] was used, together with projector augmented wave pseudo potentials[49] and a plane-wave expansion for the wave functions[50] with a cutoff energy of 600 eV. All the systems were optimized until the forces acting on each atom were smaller than 0.001 eV/Å. Since the unit cell is large consisting of 54 atoms plus the guest molecules, only the Γ point was sampled.

The binding energy between a molecule "a" and the MOF is calculated using the following equation:

$$E_B = (E_{sys-a} + E_a) - E_{sys+a}, \quad (1)$$

where $E_{sys+a}$ is the energy of the system with the "a" molecule adsorbed. $E_{sys-a}$ is the energy of the system without the adsorbed molecule and $E_a$ is the energy of the isolated "a" molecule. Note that with this definition, a *positive* binding energy corresponds to favorable binding of the molecule to the MOF.

## 3. Results.

IR spectroscopic measurements of $CO_2$ adsorption at 80 Torr are hindered by the fact that $CO_2$ gas phase absorption is too strong (no transmission signal at resonance as shown in Figure S4). Adsorbed $CO_2$ can be only reliably detected after evacuation of the gas phase. However, we show in section 3.1 that the decrease of $CO_2$ concentration within the MOFs right after evacuation is slow enough to study the effect of post-loaded guest molecule ($H_2O$, $NH_3$, $SO_2$, NO, $NO_2$ or $N_2$, $O_2$, $CH_4$) on pre-adsorbed $CO_2$. For instance, section 3.2 shows that water molecules can diffuse into the frameworks and gradually displace the pre-existing $CO_2$ trapped inside the pores (described as $H_2O \rightarrow CO_2$). In section 3.3, the co-adsorption of other molecules is considered, ($SO_2$, NO, $NO_2) \rightarrow CO_2$. The results in section 3.3 show that despite the stronger adsorption energy of the molecules ($SO_2$, NO, $NO_2$) with the metal centers, they are not able to efficiently displace the $CO_2$ molecules from the metal centers. The reason for this surprising result is examined in later in discussion section by performing first-principles calculations on the co-adsorption of $CO_2$ with $SO_2$, $H_2O$ and $NH_3$ and noting that the associated exchange barrier of $H_2O \rightarrow CO_2$ is lower than that of $SO_2 \rightarrow CO_2$. For completeness, comparative experiments of weakly bound molecules ($N_2$, $O_2$, $CH_4) \rightarrow CO_2$ are also reported in supporting information (section 12).

### 3.1. Pure $CO_2$ adsorption

The activated MOF samples were first loaded with $CO_2$ at 80 Torr for ~20 min. At this pressure, the occupation of $CO_2$ is around ~0.7 and ~0.4 molecule per metal sites for Mg-MOF-74 and Co-MOF-74, respectively.[16] The cell was then evacuated and the spectra were recorded as a function of time and summarized in Figure 1, showing a clear $CO_2$ stretching band 2350 cm$^{-1}$ after loading $CO_2$ at 80 Torr for ~20 min. It was shown in a previous study that $CO_2$ is adsorbed at the metal site, characterized by an asymmetric stretching mode $v_3$ at 2350 cm$^{-1}$, with partial occupation of the metal sites at low loading.[25] In addition to the peak at 2350 cm$^{-1}$, another shoulder is also identified at 2343 cm$^{-1}$, but quickly disappears during evacuation (i.e., less strongly bound). This band was previously assigned to $CO_2$ at the secondary sites (closest to the ligand oxygen atoms), which is consistent with observations by other techniques such as neutron powder diffraction and nuclear magnetic resonance (NMR).[25,29,31] The differential spectrum for Mg-MOF-74 in Figure 1 shows that the intensity of the $CO_2$ band at 2343 cm$^{-1}$ drops faster than the species at 2350 cm$^{-1}$, indicating that this species is adsorbed weaker than that on the metal sites.

To confirm this peak assignment, Mg-MOF-74 was first loaded with the more strongly bound species ($H_2O$, 8 Torr; $SO_2$, 150 Torr; $NH_3$, 50 Torr) to occupy the metal sites. Afterwards, the pre-loaded sample was exposed to 80 Torr $CO_2$. The binding energies of these molecules ($H_2O$, $SO_2$, $NH_3$) at the metal sites are high,[35,37] suggesting that they should remain bound at the metal site even upon introduction of $CO_2$ molecules. Hence, post-dosed $CO_2$ molecules should not be able to displace them from the metal sites and are expected to adsorb on the secondary sites instead in a sequential exposure. The observation that the post-loaded $CO_2$ is characterized by a frequency around 2342(4) cm$^{-1}$ (Figure 1b) confirms that it is adsorbed at the secondary sites instead of the metal sites. This frequency shift is in excellent agreement with the calculated shift (-7 cm$^{-1}$) that the $CO_2$ molecule undergoes as it lea-ves the metal center and moves to the center of the channel in an empty Mg-MOF-74. Furthermore the results of Figure 1b are generally consistent with the calculated values (-9 and -12 $cm^{-1}$) in $H_2O \rightarrow CO_2$ co-adsorption when $CO_2$ is displaced to the secondary sites, as discussed in Table S4 of the supporting information. The inset in Figure 1 shows that the intensity of $\nu_3$ band slowly decreases during the desorption process.

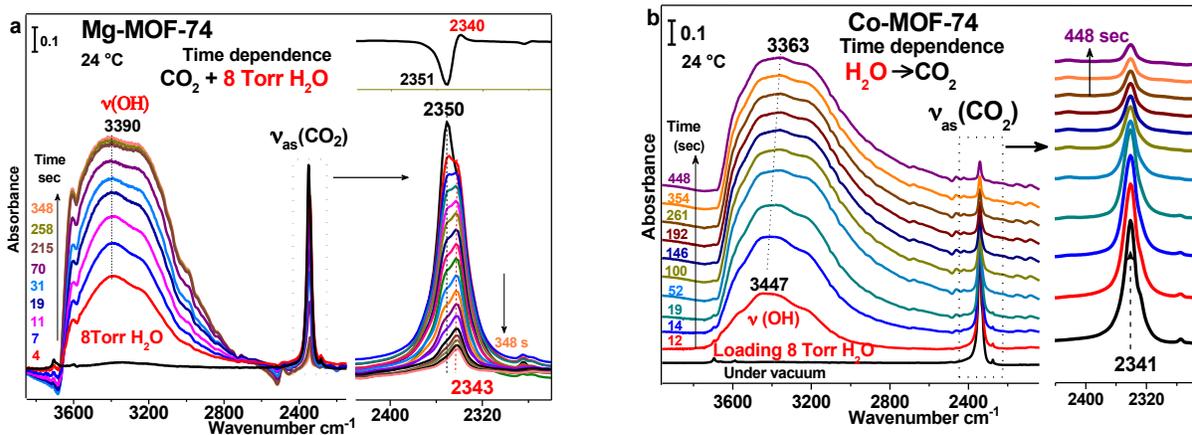

**Figure 2.** IR absorption spectra of 8 Torr $H_2O$ adsorbed into (a) Mg-, (b) Co-MOF-74 with pre-loadeded $CO_2$ as a function of time. All referenced to the activated MOF under vacuum (< 20 mTorr). The black line in the bottom shows the absorption spectra of $CO_2$ before loading water. For Mg-MOF-74, the $CO_2$ spectra contain more time points. The differential spectrum on the top right of Figure 2a shows the spectra recorded after introducing $H_2O$ for 4 s, referenced to the spectra before loading $H_2O$ as shown in the bottom black spectra. (Note that the differential peak position is slightly shifted from 2350 to 2343 $cm^{-1}$ values).

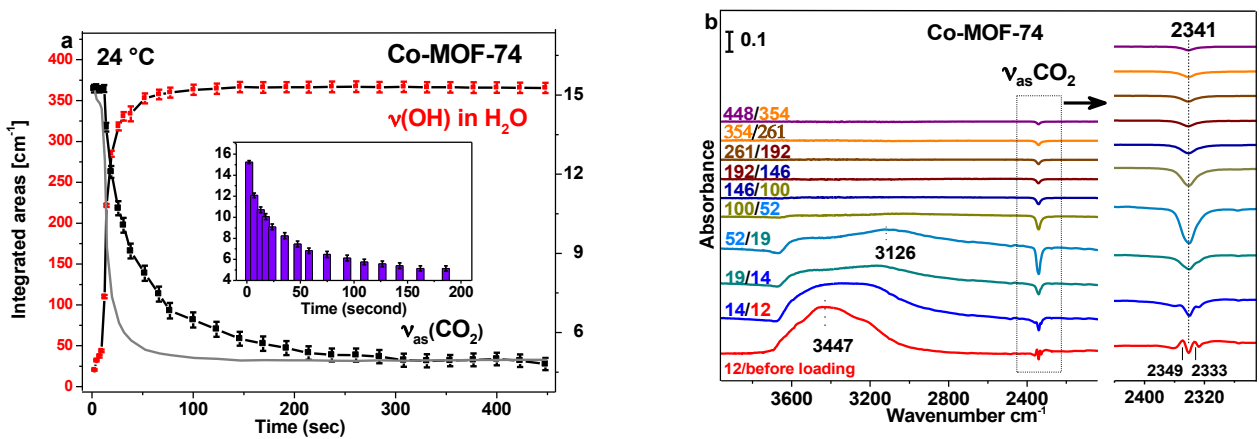

**Figure 3**. (a) Evolution of integrated areas of the IR bands of water $\nu(OH)$, red and $\nu_{as}(CO_2)$, black in Co-MOF-74 at 24 °C. The curve of $CO_2$ band is the net change due to displacement by $H_2O$ molecules after correcting for the slow release of $CO_2$ in vacuum (same for Figure 5b). The grey curve represents the expected $\nu_{as}(CO_2)$ band intensity assuming immediate water-$CO_2$ exchange (negligible exchange barrier). The inset purple bar curve shows the $\nu_{as}(CO_2)$ band evolution corrected by removing the water transport time (taken from the grey curve), hence provides the intensity evolution solely controlled by exchange kinetics. (b) Differential absorbance spectra showing the water exchange $CO_2$ process within 450 sec. Each spectrum is referenced to the previous recorded data. The color matches that of the IR absorption in Figure 2b.

Consistent with our previous measurements for Co-MOF-74, the $CO_2$ peak is found at 2341 $cm^{-1}$.[51] Compared to $CO_2$ adsorption in Mg-MOF-74, the $CO_2$ band in Co-MOF-74 is more symmetrical, indicating a more homogenous distribution of adsorbed species (see Figure S5). Upon evacuation, the peak becomes sharper, suggesting that the earlier peak broadening is related to $CO_2$-$CO_2$ vibrational coupling at high loading.[25] The shoulder at 2330 $cm^{-1}$ is assigned to the combination band of $\nu_3+\nu_2{'}-\nu_2{''}$.[52-55] The presence of trace amount of $C^{13}O_2$ is noted in the spectra with a peak at 2275 $cm^{-1}$ and 2284 $cm^{-1}$ in Mg and Co-MOF-74, respectively.

For co-adsorption studies, we use the fact that the $CO_2$ desorption rate upon evacuation is slow enough to study and compare the effects of other gases on the $CO_2$ concentration. This was done by pre-loading $CO_2$ in MOF, briefly evacuating, post-loading of another guest molecule into the MOF, and immediately monitoring the $CO_2$ desorption rates after loading the guest molecule.

### 3.2. Co-adsorption of $CO_2$ and $H_2O$

Before addressing co-adsorption, it is important to examine the interaction of pure water molecules within M-MOF-74 (M= Mg, Co, Ni). This MOF is known to be hydrophilic with high affinity to water vapor (with the adsorption heat over 60 kJ/mol[24,35]), leading to occupation of all metal sites at or above 20 % humidity level at room temperature.[56] The highest bindi-ng energy site for water molecules is the open metal site with bonding believed to occur via its oxygen atom.[57] This was confirmed spectroscopically with the detection of two sharp peaks at 3663 and 3576 cm$^{-1}$ for Mg-MOF-74 and at 3668 and 3582 cm$^{-1}$ for Co-MOF-74 at low water vapor exposure (~30 mTorr) as shown in Figure S6. The frequencies in the ranges 3660-3700 cm$^{-1}$ and 3570-3600 cm$^{-1}$ are respectively assigned to the asymmetric ($\nu_{as}$) and symmetric and ($\nu_s$) stretching modes of water molecules coordinated to the metal sites by oxygen and free from hydrogen bonding with other water molecules.[58,59] Broad bands centered at 3380, 3223 cm$^{-1}$ for Mg-MOF-74 and at 3462, 3208 cm$^{-1}$ for Co-MOF-74 intensifies with increasing water loading. It is associated with hydrogen-bonded water molecules trapped within the MOF channel due to a higher local density.

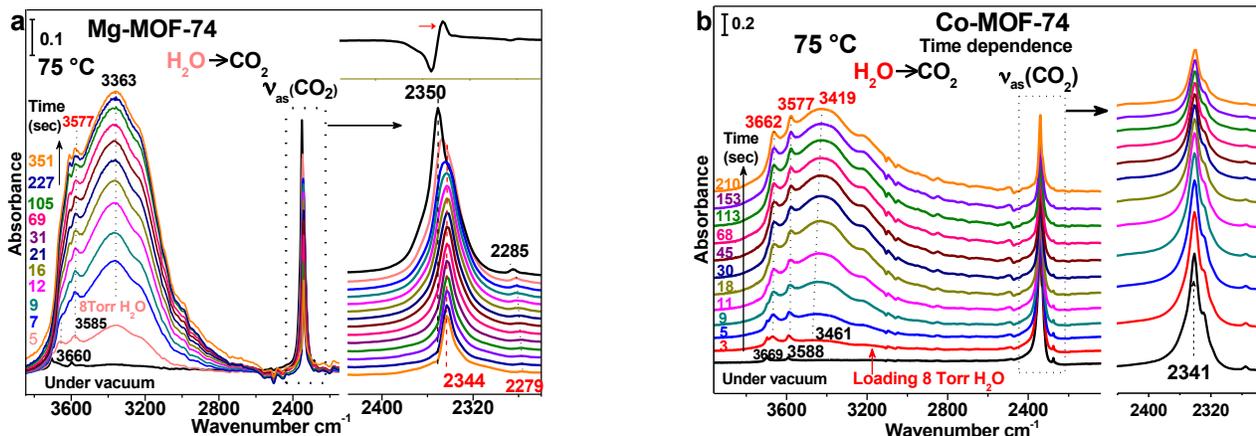

**Figure 4**. IR absorption spectra of (a) Mg-, (b) Co-MOF-74 with pre-loaded $CO_2$ exposing to 8 Torr $H_2O$ vapor as a function of time. All the spectra reference to the activated MOF in vacuum. The black line in the bottom shows the absorption spectra of $CO_2$ before loading water. The remaining spectra shows time evolution of the water adsorption and $CO_2$ desorption process. The differential spectrum on the top right of Figure 4a shows the spectra recorded after introducing $H_2O$ for 5 s, referenced to the spectra before loading $H_2O$ as shown in the bottom black spectra (note that the differential peak position is slightly shifted from 2350 to 2344 cm$^{-1}$ values).

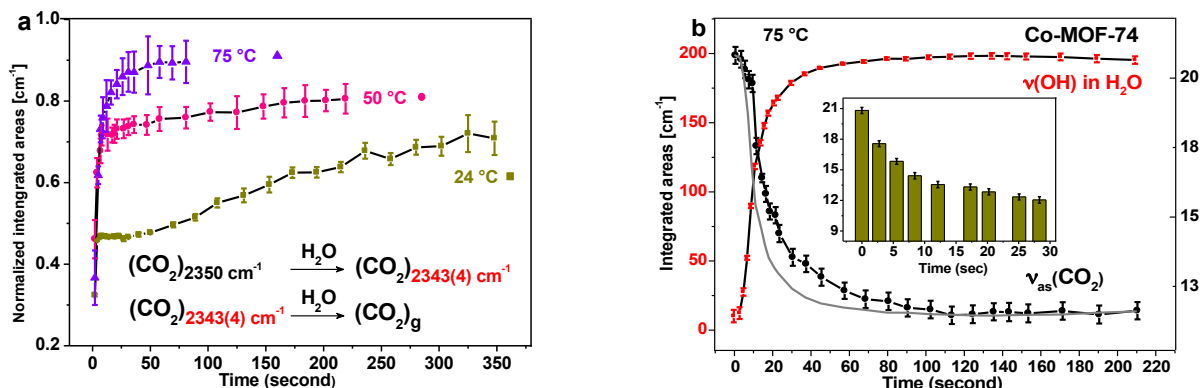

**Figure 5**. (a) Relative integrated areas of $CO_2$ band at 2343 cm$^{-1}$ (corresponding to 2344 cm$^{-1}$ band at 75 °C) as a function of time during the water adsorption into Mg-MOF-74 at 24 °C, 50 °C and 75 °C. (b) Evolution of integrated areas of the IR bands of water $\nu(OH)$, red and $\nu_{as}(CO_2)$, black in Co-MOF-74 at 75 °C. The grey curve represents the expected $\nu_{as}(CO_2)$ band intensity assuming immediate water-$CO_2$ exchange (negligible exchange barrier). The inset brown bar curve shows the $\nu_{as}(CO_2)$ band evolution corrected by removing the water transport time (taken from the grey curve), hence provides the intensity evolution solely controlled by exchange kinetics.

The $H_2O \rightarrow CO_2$ experiments were carried out by pre-loading the MOF with $CO_2$ at 80 Torr for ~20 min and measuring the $CO_2$ concentration as a function of time (i.e., desorption rate) after post-loading with $H_2O$ molecules. Activated and $CO_2$-loaded Mg-MOF-74 were exposed to 8 Torr $H_2O$ after evacuation of gas phase $CO_2$ for ~7 min (i.e., after the cell pressure dropped below 200 mTorr and the gas $CO_2$ signal was negligible). A pressure of 8 Torr corresponds to a humidity level of 40% at ambient temperature, which is sufficient to fully saturate all of the open metal sites of an empty MOF-74.[56]

Time-dependent spectra in Figure 2, Figure S9, and Figure S10 show that exposing MOFs to water vapor results in a significant drop in the intensity of $\nu_{as}(CO_2)$ band for both Mg- and Co-MOF-74. Correspondingly, the intensity of the water stretching bands $\nu(OH)$ rapidly increases and saturates. Figure 3a and Figure S11a show the transport process of water molecules within MOFs pellet. At room temperature, an equilibrium is reached after ~350 sec for Mg-MOF-74 and ~150 sec for Co-MOF-74. Clearly water is gradually adsorbed and exchanged with pre-adsorbed $CO_2$ after 8 Torr vapor exposure. For Mg-MOF-74, the peak of the $CO_2$ stretching band shifts from 2350 cm$^{-1}$ to 2343 cm$^{-1}$ (see the differential spectrum in the top portion of Figure 2a), indicating that the water molecules displace the pre-adsorbed $CO_2$ molecules to either secondary sites (closest to the ligand oxygen atoms) or into the channel. With more water molecules adsorbed into Mg-MOF-74, the intensities of both $CO_2$ bands at 2350 and 2343 cm$^{-1}$ decrease. For Co-MOF-74, this is also shown by plotting the intensity of the two bands $\nu_{as}(CO_2)$ and $\nu(OH)$ as a function of time and comparing the $CO_2$ desorption and water adsorption rates. The error bars for the integrated areas shown in Figure 3a are mostly due to uncertainties associated with baseline determination. Since the $CO_2$ intensity slowly decreases even in the absence of another guest molecule (see Figure S10a), the $CO_2$ integrated area (see Figure S10a) is used as reference to take into account the removal rate that would occur in vacuum independent of water. In Figure 3a, the measured integrated intensities of both the $\nu_{as}(CO_2)$ and $\nu(OH)$ modes are plotted in black and red, respectively. The black curve in Figure 3a shows the evolution of $CO_2$ band as water diffuses within MOF crystal. The $CO_2$ band intensity decreases more slowly than the water band intensity increases. As a result, after the water adsorption reaches equilibrium (~150 sec), while the $CO_2$ concentration continues to decrease for another ~300 sec. This highlights the slower removal of pre-existing $CO_2$ molecules from the metal sites in the MOFs. In order to take into account the time that takes for the water molecule to diffuse through pores inside the crystal (transport delay), the expected evolution of $CO_2$ assuming that immediate exchange takes place (i.e. negligible exchange barrier) is plotted in solid grey line in Figure 3a by taking water uptake complement (red curve in Figure 3a) (grey curve). Clearly, the black data points do not overlap this grey curve, suggesting that additional time is required for exchange (i.e. the barrier is not negligible). The normalized data, correcting for the time due to transport (taken from the grey curve) are shown in the inset of Figure 3a. Also, the differential spectra in Figure 3b clearly shows this delayed removal or exchange between the water molecules and preadsorbed $CO_2$ in the framework. The $CO_2$ stretching bands are perturbed, characterized by the shift of the differential spectrum (2349 cm$^{-1}$, 2333 cm$^{-1}$) in the very beginning of loading water, i.e., within 12 sec. Later on, the OH mode red shifts from 3447 to 3363 cm$^{-1}$ (Figure 3b) due to H-bonding, and its intensity increases when more and more water molecules are adsorbed into MOFs. After ~150 sec, water adsorption gradually reaches equilibrium and the intensity of $CO_2$ bands continues to drop, indicated by the loss (negative) band at 2341 cm$^{-1}$ in Figure 3b.

To determine if the exchange process of $CO_2$ by $H_2O$ is activated, experiments were carried out at higher temperatures (50 °C in Figure S12 and 75 °C in Figure 4). It was not feasible to go to temperatures above 100 °C because $CO_2$ would desorb too quickly from the framework. MOF film thicknesses on the KBr pellet used in temperature dependent experiment are similar in each single measurement (see Figure S13 and Figure S14). It is therefore possible to compare the equilibrium time at different temperatures in a quantitative way. After loading $CO_2$ and evacuating the gas phase for ~5 min (base pressure < 200 mTorr), 8 Torr water was introduced into the cell and the spectra were recorded as function of time, as shown in Figure 4 for both Mg- and Co-MOF-74. The water stretching band above 3000 cm$^{-1}$ grows and saturates after ~210 sec for Mg-MOF-74 and ~90 sec for Co-MOF-74, i.e., faster than at room temperature (Figure 3a and Figure S11). The presence of sharp modes at 3660, 3577 cm$^{-1}$ in the spectra of Mg-MOF-74 and at 3662, 3577 cm$^{-1}$ in the spectra Co-MOF-74 indicates that there are fewer hydrogen-bonded "bulk" water molecules accumulated inside the channel than at room temperature. The red shifts of the water stretching bands, including the sharp bands in Mg-MOF-74 from 3585 cm$^{-1}$ to 3577 cm$^{-1}$ and the broad band in Co-MOF-74 from 3461 cm$^{-1}$ to 3419 cm$^{-1}$, are all consistent with the population of sites adjacent to the metal sites with increasing hydrogen bonding because of intermolecular interaction.[25]

The intensity of the $CO_2$ stretching band decreases once water vapor is introduced. A similar shift of the band from 2350 cm$^{-1}$ to 2344 cm$^{-1}$ was also observed for $CO_2$ in Mg-MOF-74 at 75 °C, as confirmed in the differential spectrum on top of Figure 4a. Note that at 75 °C, the peak positions are now at 2350 and 2344 cm$^{-1}$ for $CO_2$ adsorbed in Mg-MOF-74 at the primary (metal) sites and secondary sites. The stretching band of the isotopic $C^{13}O_2$ species also shifts from 2285 to 2279 cm$^{-1}$, confirming that $CO_2$ is displaced by $H_2O$. It is apparent that the ratio of the 2344 cm$^{-1}$ band (corresponding to 2343 cm$^{-1}$ band at room temperature) to the 2350 cm$^{-1}$ band intensities is larger than the corresponding ratio at room temperature (see Figure 2a), indicating that more molecules with 2350 cm$^{-1}$ band are transformed into molecules with 2344 cm$^{-1}$ band at higher temperatures. At the same time, $CO_2$ at the secondary sites diffuses out of the framework due to adsorption of additional water molecules.

The initial displacement of $CO_2$ by $H_2O$ in Mg-MOF-74 is clearly reflected by the transformation of 2350 cm$^{-1}$ band to 2343(4) cm$^{-1}$ (i.e. 2344 cm$^{-1}$ at 75 °C) band. Then the overall intensity decreases upon loading water. To evaluate this initial $CO_2$ displacement as a function of water loading and temperature, the total $CO_2$ bands are fitted into two band (see Figure S12) and the relative intensity (normalized to the total $CO_2$ intensity at each point) of the $CO_2$ band at 2343(4) cm$^{-1}$ is plotted as function of time for different temperatures (24 °C, 50 °C, 75 °C) in Figure 5a. The scheme shown in this figure illustrates the process of the two-steps $CO_2$ evacuation from Mg-MOF-74. The time dependent curves in Figure 5a show that a larger fraction of $CO_2$ molecules with a 2350 cm$^{-1}$ band is transformed into molecules with a 2343 cm$^{-1}$ band (corresponding to 2344 cm$^{-1}$ band at 75 °C) as more water is adsorbed into Mg-MOF-74. It is apparent that water inclusion into the frameworks causes this intensity shift from 2350 cm$^{-1}$ to 2343 cm$^{-1}$. The transformation proceeds and equilibrates faster at 50 °C (~200 sec), 75 °C (~75 sec) since the diffusing of water molecules within Mg-MOF-74 is sped up (see Figure S11). We also note that the 2343 cm$^{-1}$ band is more intense for high-temperature $H_2O$ exposures (relative intensity: ~0.9 for 75 °C and ~0.8 for 50 °C), which suggests that increasing the temperature promotes the overall intensity shift of the band from 2350 cm$^{-1}$ to 2343 cm$^{-1}$. Figure 5b shows the evolution of integrated areas of both water and $CO_2$ stretching modes over ~3.5 min in Co-MOF-74. The inset shows the exchange process after correcting the time due to transport. Compared to the results at room temperature shown in Figure 3a, two main observations are noted: (1) it takes less time (~90 sec) for water molecules to reach equilibrium in Co-MOF-74; (2) the desorption rate of $CO_2$ is faster and closer to that of water adsorption rate; specifically, the difference between the measured and corrected dependences of the 2341 cm$^{-1}$ band at 75 °C (Figure 5b) is smaller than the one at 24 °C (see Figure 3a), i.e. the displacement of $CO_2$ by water is faster at 75 °C than at room temperature.; the $CO_2$ stretching band intensity decreases by ~50 % within ~120 sec after loading $H_2O$. The remain-ing $CO_2$ coexists with the loaded water molecules.

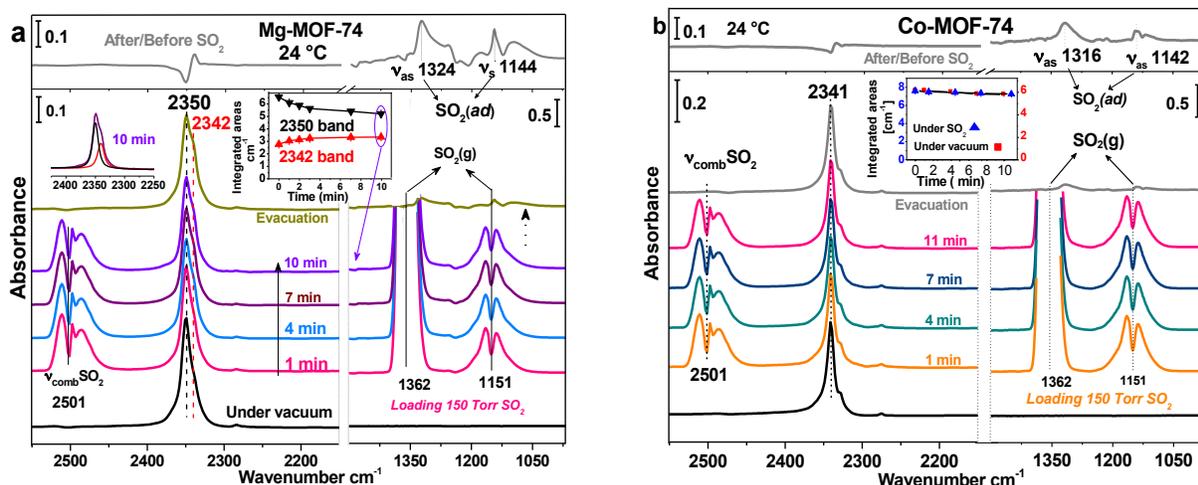

**Figure 6**. IR absorption spectra of activated and $CO_2$–preloaded (a) Mg-, (b) Co-MOF-74 exposed to 150 Torr $SO_2$. The bottom spectra show the adsorbed $CO_2$ in Mg, Co-MOF-74 under vacuum (< 20 mTorr). The middle part shows the time-dependent spectra during the $SO_2$ exposure and the spectra after evacuation of gas phase $SO_2$. The differential spectra on the top right of Figure 6 (a) and (b) shows the spectra recorded after loading $SO_2$ for 10 min and evacuation of gas phase, referenced to the spectra before loading $SO_2$ as shown in the bottom black spectra (note that a shoulder is growing at 2342 cm$^{-1}$).

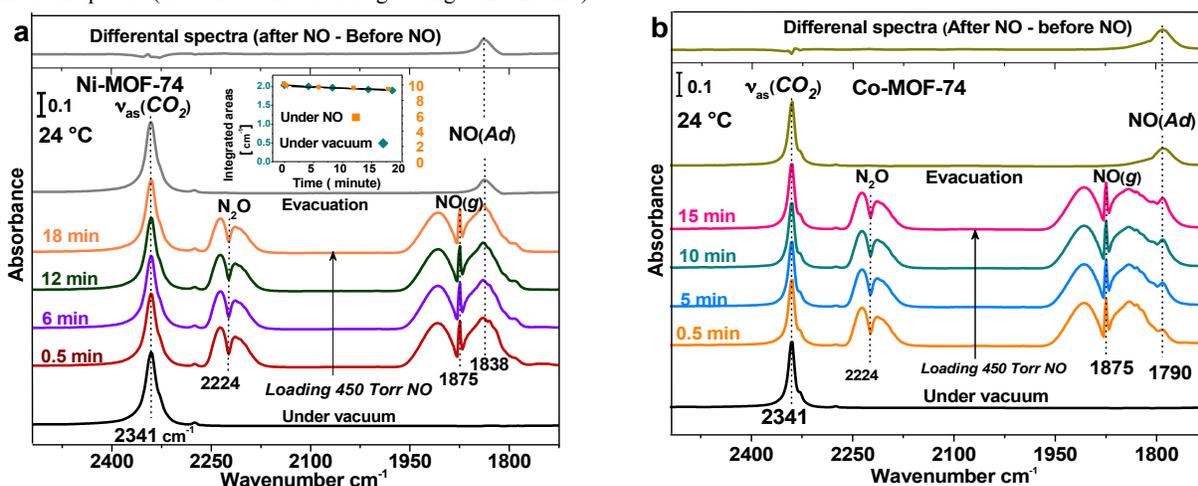

**Figure 7.** (a) IR absorption spectra of activated Ni-MOF-74 with pre-adsorbed $CO_2$ exposed to 450 Torr NO. Color scheme: black line, before loading NO; the top grey line, after evacuation of gas phase NO for 20 sec. (b) IR absorption spectra of activated and $CO_2$– preloaded Co-MOF-74 exposed to 450 Torr NO. Color scheme: black line, before loading NO; dark yellow, after evacuation of gas phase for 20 sec. The other colors correspond to time-dependent spectra under the $NO_2$ gas phase, as indicated. All spectra are referenced to the activated MOF in vacuum except the differential spectra shown on the top, which is obtained by subtracting the spectra before 450 Torr NO exposure from the spectra after evacuation 450 Torr NO gas phase.

The main finding for this section is that water can readily diffuse into the framework of M-MOF-74 (M= Mg, Co) and displace the pre-loaded $CO_2$. Furthermore, there is an observable kinetic barrier controlling the molecular exchange process of $CO_2$ by $H_2O$.

### 3.3. Co-adsorption of $CO_2$ and $SO_2$, NO, $NO_2$

Similar experiments were performed to determine the behavior of acid gases such as $SO_2$, NO, $NO_2$ when these are co-adsorbed with $CO_2$ in M-MOF-74 (M= Mg, Co, Ni). Previous breakthrough measurements have shown that Mg-MOF-74 has the highest adsorption capacity for $SO_2$ compared to M-MOF-74 (M= Zn, Ni and Co).[56] Moreover, DFT calculations have shown that $SO_2$ binds more strongly (>70 kJ/mol) to the open metal sites ($Mg^{2+}$) in Mg-MOF-74 than $CO_2$ molecules (~47 kJ/mol).[35,36,60] Co-MOF-74 ranks second for $SO_2$ uptake and the $SO_2$ binding energy is slightly less than in Mg-MOF-74.[27,36] Nitric oxide (NO) molecules are also strongly bound to the open metal sites ($Ni^{2+}$ and $Co^{2+}$) of Ni- and Co-MOF-74, forming a 1:1 nitrosyl adduct with a large heat of adsorption (~90 kJ/mol) in Ni-MOF-74.[38,61] The interaction of NO with Mg-MOF-74 is the weakest (33 kJ/mol), mostly due to the absence of d-electrons that prevent back bonding.[35] Therefore we chose to study the $SO_2 \rightarrow CO_2$ process in M-MOF-74 (M= Mg, Co) and the $NO \rightarrow CO_2$ process in M-MOF-74 (M= Ni, Co).

The IR absorption measurements were performed by introducing $SO_2$ gas into M-MOF-74 (M= Mg, Co) and NO gas into M-MOF-74 (M= Ni, Co), both pre-loaded with $CO_2$. Since the concentrations of $SO_2$ and $NO_x$ (NO, $NO_2$) in flue gas are 800 ppm and 500 ppm, respectively,[3] we purposely introduced much higher pressures (~ 150 Torr for $SO_2$ and ~ 450 Torr for NO) to test the stability of $CO_2$ under enhanced competitive adsorption conditions at high pressure. According to Bonino and McKinlay's isotherm measurements,[38] such a high pressure of NO is enough to saturate all the available metal sites in activated (i.e. empty) MOF-74. $SO_2$ isotherm measurements[62] in Mg-MOF-74 show that the uptake of ~7.5 mmol/g at 150 Torr, corresponds to ~0.91 $SO_2$ molecule per open metal site.

The activated sample of Mg-MOF-74 was first exposed to 80 Torr of $CO_2$ and then evacuated for 1 h until the base pressure reaches below 20 mTorr. This procedure is different from what was used for water co-adsorption because the $SO_2$ replacement process is slow (as described in the following part). Therefore, $SO_2$ was introduced into the cell only after evacuation for 1 h. Figure 1 shows that, after 1 h pre-evacuation, the remaining $CO_2$ concentration only decreases by <5 % within ~20 min. The slow decrease rate allows us to characterize the effect of $SO_2$ loading on the pre-adsorbed $CO_2$. Considering that adsorption of trace amount of water impurity out-gassing from the cell will contribute to $CO_2$ desorption, especially for Mg-MOF-74 that has a large affinity to $H_2O$ molecules, the activated M-MOF-74 (M= Mg, Co) pre-loaded with $CO_2$ was exposed to $SO_2$ for ~10 min. Figure 6 shows that the gas phase $SO_2$ asymmetric stretching bands dominate in the IR spectra (their absorbance is off scale in the region of 1000 to 1500 cm$^{-1}$). The broad peak centered at 2501 cm$^{-1}$ is the $SO_2$ combination band ($\nu_{as}+\nu_s$). The diffusion of $SO_2$ molecules within MOFs is measured by monitoring the symmetric stretching band $\nu_s(SO_2)$ as a function of time after subtraction of gas phase spectra. As shown in Figure S16, $SO_2$ adsorption reaches equilibrium in both Mg- and Co-MOF-74 within ~10 min.

The intensity of the $CO_2$ band in Figure 6a does not vary significantly, certainly not close to the variations observed in $H_2O \rightarrow CO_2$ experiments (Figure 2), except for a notable growth of a shoulder at 2342 cm$^{-1}$ during the $SO_2$ loading in Mg-MOF-74. The differential spectrum on top of Figure 6a highlights the transformation of some species with a band at 2350 cm$^{-1}$ into species with a band at 2342 cm$^{-1}$. The integrated areas of the 2350 cm$^{-1}$ and 2342 cm$^{-1}$ components are plotted as a function of time and summarized in the inset in Figure 6a. The band at 2350 cm$^{-1}$ decreases by ~20% and the band at 2342 cm$^{-1}$ increases by ~20% after ~10 min. The adsorption of $SO_2$ in Mg-MOF-74 is monitored by detecting two bands at 1324 and 1144 cm$^{-1}$, assigned to the asymmetric ($\nu_{as}$) and symmetric ($\nu_s$) stretching bands of $SO_2$, respectively, after evacuation of the gas phase.[63-65] For Co-MOF-74, the perturbation of the $CO_2$ band induced by $SO_2$ co-adsorption is much smaller with only a very small red shift as shown in the top differential spectrum of Figure 6b, which indicates that no significant changes occur upon $SO_2$ loading. The inset in Figure 6b shows that the $CO_2$ loss rate of (band at 2341 cm$^{-1}$) upon $SO_2$ loading is similar to that under evacuation in vacuum (< 20 mTorr), which is clearly different from $H_2O \rightarrow CO_2$ in Co-MOF-74. After ~11 min, the intensity drops by less than 5%, and the presence of adsorbed $SO_2$ can be observed at 1316 and 1142 cm$^{-1}$ in the Co-MOF-74 spectra.

For Ni-MOF-74, the $CO_2$ band observed at 2341 cm$^{-1}$ after $CO_2$ loading at 80 Torr and subsequent evacuation are similarly strong, which is consistent with a previous study.[28] After the sample with pre-loaded $CO_2$ was evacuated for 1 h and residual pressure was reduced below 20 mTorr, the sample was exposed to 450 Torr NO gas and the spectra were recorded for over ~15 min. This is enough for NO uptake to reach equilibrium as recorded by monitoring the NO stretching band absorption as a function of time (Figure S17). The isotherm measurements show that most of the metal sites could be occupied at this pressure.[61] The gas phase NO stretching band is centered at 1875 cm$^{-1}$ and the peak at 1838 cm$^{-1}$ is due to NO coordinated to the metal sites.[38] The band centered at 2224 cm$^{-1}$ is due to the stretching mode of $N_2O$ that co-exists with NO in equilibrium.[66] After evacuation of gas phase NO, adsorbed NO characterized by an absorption band at 1838 cm$^{-1}$ can be clearly seen in IR spectra (see Figure 7a). The inset in Figure 7a summarizes the time dependence of the $CO_2$ band intensity at 450 Torr of NO, compared to the $CO_2$ band under vacuum before introducing NO. After NO gas exposure, the $CO_2$ stretching band only loses ~5 % of its intensity within ~20 min, which is similar to the $CO_2$ desorption rate in vacuum (< 20 mTorr) under the same conditions (see inset of Figure 7a and Figure S18). This is totally different from $H_2O$ exposure, for which the $CO_2$ intensity drops by ~90% (see Figure S19). The differential spectrum is obtained by subtracting the spectra before NO exposure (see the black spectra in the bottom in Figure 7a) from those obtained after 450 Torr NO exposure for ~18 min and evacuation (see the grey spectra after evacuation of gas phase for 20 sec in Figure 7a). This line shows only a slight change in $CO_2$ absorption band at 2341 cm$^{-1}$. The amount of adsorbed NO is less by ~40% in the sample with pre-adsorbed $CO_2$ in Ni-MOF-74, compared to the adsorption in the empty sample (see Figure S20). The same experiment was repeated in Co-MOF-74 by exposing MOFs to 450 Torr NO and recording spectra as a function of time, see Figure 7b. From this figure it can be seen that the adsorbed NO appears at 1790 cm$^{-1}$ and the $CO_2$ band is only slightly affected by NO loading as indicated by the differential spectrum. It should be mentioned that this differential spectrum was obtained by subtracting the spectra before NO exposure from those obtained after 450 Torr NO exposure for ~15 min and evacuation. Figure 7b also shows that the $CO_2$ stretching band decreases by less than ~5% within ~15 min, which is similar to the concentration of upon $SO_2$ loading shown in Figure 6b. This observations indicate that NO has no effect on the pre-adsorbed $CO_2$.

In addition to $SO_2$ and NO, we also examined the $NO_2 \rightarrow CO_2$ process in M-MOF-74 (M= Mg, Co), as reported in supporting information. The case of $NO_2$ is interesting because $NO_2$ is both molecularly and dissociatively adsorbed into M-MOF-74 (M= Zn, Mg, Co, Ni), forming NO, $NO_3^-$, and adsorbed $NO_2$ species (see Figure S21). Performing a similar experiment for $NO_2 \rightarrow CO_2$ in M-MOF-74 (M= Mg, Co), as described in Figure S21, we also found that $NO_2$ dissociates but the $CO_2$ absorption band $\nu_{as}(CO_2)$ remains mostly unaffected in both Mg- and Co-MOF-74 after ~10 min of $NO_2$ exposure.

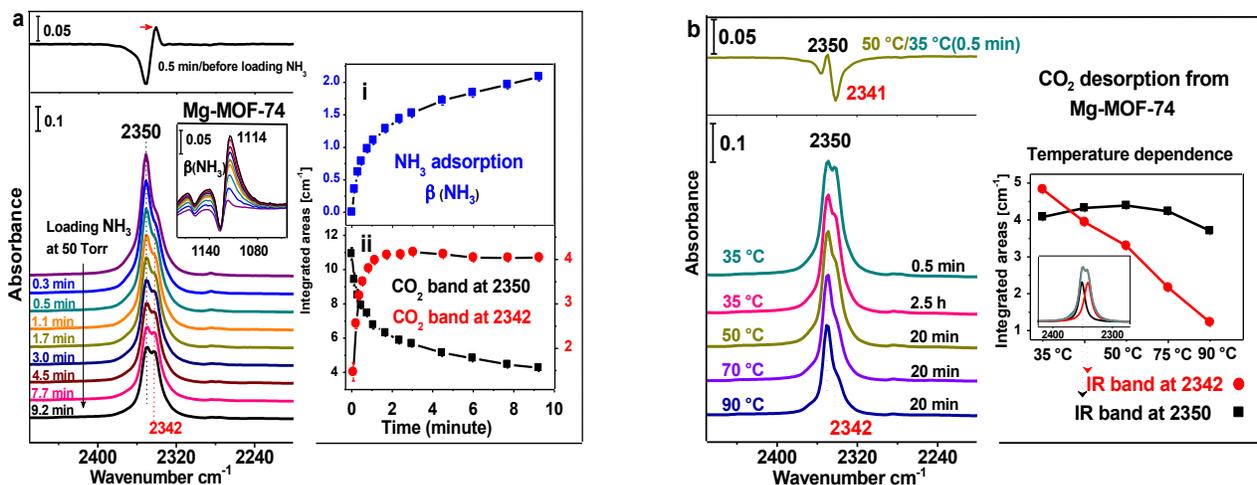

**Figure 8.** (a) Time-dependent IR absorption spectra of activated and $CO_2$ – preloaded Mg-MOF-74 after exposure to 50 Torr $NH_3$. The top is the spectrum (after loading $NH_3$ for 0.5 min) referenced to the MOF spectrum before $NH_3$ loading. Note that the differential peak position is slightly shifted from 2350 to 2342 cm$^{-1}$. The top-inset shows that the $NH_3$ bending mode increases and the bottom inset shows the evolution of the two components at 2350 cm$^{-1}$ and 2342 cm$^{-1}$ as a function of time (fitting is shown in Figure 8b). The spectra of the bending mode $\beta(NH_3)$ region are shown in the middle inset. (b) Temperature-dependent IR spectra of $CO_2$ desorption from Mg-MOF-74. The data was recorded after evacuation at 35 °C for 2.5 h and then at 50 °C, 70 °C and 90 °C for 20 min. The right plot shows that the two components at 2350 and 2342 cm$^{-1}$ change as a function of temperature. All spectra are referenced to the activated MOFs.

### 3.4. Co-adsorption of $CO_2$ and $NH_3$

In a similar manner to what was done with the molecules discussed above, we further tested the co-adsorption of $CO_2$ with $NH_3$ to compare with $H_2O$ and provide quantitative data for *ab initio* calculation.

Before introducing $NH_3$ molecules, the gas phase $CO_2$ was evacuated for ~7 min until the pressure dropped below 200 mTorr and the primarily adsorbed $CO_2$ can be seen at 2350 cm$^{-1}$ in Mg-MOF-74 (see Figure 8a). After exposing the activated Mg-MOF-74 to 50 Torr $NH_3$, the increase of adsorbed $NH_3$ can be seen by the strengthening of its bending mode $\beta(NH_3)$ at 1114 cm$^{-1}$ (the spectra in the middle-inset and intensity in inset (i) of Figure 8a).[67] The intensity of the $CO_2$ stretching band decreases once $NH_3$ adsorbs into the MOF (see Figure 8a and Figure S22). Similar to the $H_2O \rightarrow CO_2$ measurements summarized in Figure 2, the band centered at 2342 cm$^{-1}$ becomes more intense during $NH_3$ loading. The integrated areas of this band increases fast in the beginning and stabilize after ~3 min; while the band at 2350 cm$^{-1}$ continually decreases as shown in the spectra in Figure 8a and the inset (ii) of this figure. This observation suggests that the species at 2350 cm$^{-1}$ is first converted into the species at 2342 cm$^{-1}$ by $NH_3$ loading and then desorbed from the frameworks. The conversion is more clearly indicated by the differential spectrum in Figure 8a (plotted by referencing the spectrum after 0.5 min $NH_3$ exposure to the MOF spectrum before loading $NH_3$): there is a little gain for the component at 2342 cm$^{-1}$, arising from the displacement of $CO_2$ from the metal site (2350 cm$^{-1}$) to the secondary site (2342 cm$^{-1}$). After ~3 min, an equilibrium is reached between conversion of the $CO_2$ at metal sites to the secondary site and desorption of $CO_2$ from the secondary site. After approximately 9 min exposure, the $CO_2$ band 2350 cm$^{-1}$ decreased by ~60%. Then desorption was induced by evacuation as a function of temperature to check the binding strength of these two types of adsorbed $CO_2$ molecules. As shown in Figure 8b, the $CO_2$ at 2342 cm$^{-1}$ is more readily removed than that at metal sites by the desorbed $CO_2$ from secondary sites. However after annealing at 50 °C, the 2350 cm$^{-1}$ band also decreases, indicating that refilling of $CO_2$ from the secondary sites is not enough. Although the secondary sites are easier to remove under vacuum as shown in Figure 1a and Figure 8b, they are less susceptible to be attacked by the incoming $NH_3$ molecules. $NH_3$ molecules displace the $CO_2$ adsorbed on the open metal sites first (see Figure 8a). This suggests that the driving force for displacing the $CO_2$ is the strong affinity of $NH_3$ to the metal sites $Mg^{2+}$.

The faster displacement process of $NH_3 \rightarrow CO_2$ and $H_2O \rightarrow CO_2$ suggest that molecules that can establish a hydrogen bond are more effective in removing pre-existing $CO_2$. This point is thoroughly explained and quantified by first principles calculations in Section 4.3.

### 4. Discussion
#### 4.1 Differentiating between transport and exchange kinetics in $H_2O \rightarrow CO_2$ process

For water, the time needed for adsorption to reach an equilibrium in Co-MOF-74 pellet with ~40 μm thickness (red line in Figure 3a) is temperature dependent; it requires ~150 sec at 24 °C and ~90 sec at 75 °C. This points to a finite diffusion time (transport of water molecules into the MOF structure). A diffusion pathway was predicted by the previous ab initio simulations[24], which suggested that the molecule moves longitudinally along the c axis through the channel with the lowest diffusion barrier (0.06 eV). Such an activated diffusion process needs to be taken into account in a system where $CO_2$ is pre-adsorbed to separate mass transport from other kinetic mechanisms such as exchange kinetics.

When $CO_2$ is pre-adsorbed and the cell was evacuated, its desorption is sped up by water post-loading (Figure. 2 and 4). However, the removal of $CO_2$ is strikingly slower than the uptake of water. This can be seen by taking the complement of water uptake (red curves in Figure 3a and Figure 5b) to obtain the predicted $CO_2$ release if no other kinetic barriers exist (grey curve). This grey curve represents the amount of $CO_2$ that would be displaced by water if the exchange was instantaneous (i.e. with no kinetic barrier). This simulated $CO_2$ concentration curve is below the observed $CO_2$ concentration, highlighting that the measured $CO_2$ concentration (black curve) evolved more slowly. There is therefore a kinetic barrier associated with the exchange reaction of $H_2O \rightarrow CO_2$.

The result of subtracting the water transport time (taken from the grey curve) from the $CO_2$ removal time is shown in the insets of Figure 3a and Figure 5b. In summary, the black curve shows the measured $CO_2$ concentration and the inset shows the kinetics of exchange only, after transport has been normalized out.

The analysis of $H_2O \rightarrow CO_2$ in Mg-MOF-74 is similar. However, the removal of $CO_2$ molecules within the framework proceeds in two steps: $CO_2$ molecules are initially pushed to the secondary sites by water adsorption; $CO_2$ gradually desorbs from the secondary sites and exit the framework after more water is adsorbed. While the overall intensity of the two bands 2350 cm$^{-1}$ and 2343(4) cm$^{-1}$, associated with metal and secondary sites, decreases during water loading, the relative intensity of the $CO_2$ band at 2343(4) cm$^{-1}$ (normalized to the total $CO_2$ intensity) increases as water molecules gradually diffuse into the frameworks, as $CO_2$ is first displaced to the secondary site (consistent with a finite time necessary to remove $CO_2$ from the framework through mass transport). Note that, at 50 °C and 75 °C, the diffusion of both water and $CO_2$ within the framework is higher (Figure S11), as is the exchange rate (Figure 5a).

### 4.2. Ability of guest molecules to displace $CO_2$ from the metal adsorption site in M-MOF-74 (M= Mg, Co, Ni) and relationship to binding energies

Among all tested gases $H_2O$, $NH_3$, $SO_2$, NO, $NO_2$, $N_2$, $O_2$, and $CH_4$ (see the results for $N_2$, $O_2$, $CH_4$ in Section 12 of the supporting information), only $H_2O$ and $NH_3$ molecules are observed to readily displace the pre-adsorbed $CO_2$ bonded to the metal sites. Water is known to have a large binding energy to the frameworks, especially to the open metal sites. To make the displacement reaction take place, the process should be thermodynamically favorable overall. In other words, the binding energies of the displacing molecules to the adsorption sites should be larger than those of the displaced species. Table 1 summarizes all the adsorption enthalpy of small molecules adsorption into MOF-74. Some of them ($H_2O$, $SO_2$) are taken from theoretical studies by DFT calculations and some of them ($CO_2$, NO, $N_2$, $CH_4$) are calculated from zero-coverage isosteric heat of adsorption derived by isotherm measurements. Yu's previous calculation[35] reported that the binding energies for $H_2O$ and $SO_2$ in Mg-MOF-74 are 80 and 73 kJ/mol, respectively. However, our calculation results based on the optimized structure show that $SO_2$ has a larger binding energy (~90 kJ/mol, see Table S3) in Mg-MOF-74 than $H_2O$ because of multipoint interaction of $SO_2$ with the Mg-MOF-74 structure: one oxygen atom (of $SO_2$) engages the metal site ($Mg^{2+}$) and establishes the primary interaction with MOFs, whereas the sulfur atom also interacts directly with one of the carbon atoms of the benzene ring of the linker. In contrast, the water heat of adsorption in Mg-MOF-74 is 73.2 kJ/mol.[60] The calculations for the co-adsorption system of $H_2O \rightarrow CO_2$ and $SO_2 \rightarrow CO_2$ presented in the next section validate the binding energy trend: $SO_2 > H_2O$.

On the other hand, no available data can be found in the literature for NO adsorption in Co-MOF-74 and for $NH_3$ in Mg-MOF-74. However, McKinlay and coworker's[61] isotherm measurements for NO adsorption in Co-MOF-74 showed that the desorption curve of the experiments exhibits considerable hysteresis, similar to NO desorption in Ni-MOF-74, which is consistent with a strong interaction of NO with coordinatively unsaturated $Co^{2+}$ ions. The uptake of $NH_3$ has been measured in M-MOF-74 (M= Zn, Mg, Co, Ni) by breakthrough measurements, showing that $NH_3$ can be significantly adsorbed (7.6 mol/kg) in Mg-MOF-74 for 1000 mg·m$^{-3}$ (typical of feed concentration) at room temperature.[56] Although DFT evaluates the molecular binding energy of $NO_2$ molecules to be around 41 kJ/mol, similar to $CO_2$ molecules, the experimental results show that $NO_2$ molecules are actually quite reactive with the metal centers, as evidenced by dissociation into nitric oxide and nitrate species (see Figure S21).

*Table 1.* **Summary of adsorption energies (kJ/mol) of small molecules in Mg, Ni and Co-MOF-74 compounds**

| MOF-74 Mg, Ni, Co | $CO_2$ | $H_2O$ | $SO_2$ | NO | $NO_2$ | $N_2$ | $CH_4$ |
|---|---|---|---|---|---|---|---|
| Mg | 43-48$^a$ | 73-80$^b$ | 73-90$^c$ |  | 41$^i$ | 21$^d$ | 19$^e$ |
| Ni | 41$^f$ | 61$^g$ |  | 90-92$^h$ |  |  | 20$^e$ |
| Co | 37-41$^f$ | 72$^g$ |  |  |  |  | 20$^e$ |

Note: a is taken from the reference ($^{4,35,68-70}$); b is taken from the reference ($^{35,60}$); c is take from the reference (35); d is taken from the reference ($^{70}$); e is taken from the reference ($^{71}$); f is taken from the reference ($^{4,60}$); g is taken from the reference (38); h is taken from the reference (38). i is taken from reference (35).

From Table 1, we can see that $SO_2$ and NO molecules bind significantly more strongly (by a factor of two) than $CO_2$ molecules in M-MOF-74 (M= Mg, Co, Ni), which is similar or even higher than $H_2O$ molecules. However, the exchange of $CO_2$ by $SO_2$ and NO molecules is much slower than $H_2O$. For a period between 10 to 20 min (gas uptake reaches equilibrium), only $SO_2$ can partially displace the $CO_2$ molecules in Mg-MOF-74 as shown in Figure 6a. For $SO_2$ in Co-MOF-74, only a small perturbation of $CO_2$ vibrations is detected by differential-looking features in the difference spectra, see Figure 6b. Compared to the $H_2O$ exposure in M-MOF-74 (M= Mg, Co, Ni), NO and $NO_2$ exposures only induce very small changes in the spectrum of the adsorbed $CO_2$ species after NO and $NO_2$ molecule incorporation. Even though the open metal sites are actively involved in $NO_2$ dissociation, the pre-adsorbed $CO_2$ at 2350 cm$^{-1}$ in Mg-MOF-74 and 2341 cm$^{-1}$ in Co-MOF-74 is not strongly affected by $NO_2$ adsorption. Besides, pressure dependence measurements upon $N_2$ (Figure S24), $O_2$ (Figure S25), and $CH_4$ (Figure S26) loading into Mg-MOF-74 also show that the pre-adsorbed $CO_2$ remains unaffected by these weakly adsorbed molecules.

In summary, even though $SO_2$, NO, and $NO_2$ molecules either have a larger binding energy or are more reactive in the frameworks than $CO_2$, the pre-adsorbed $CO_2$ is hardly replaced by these molecules while $H_2O$ and $NH_3$ molecules readily replace $CO_2$. From these results, we conclude that the kinetic barriers for $CO_2$ replacement (exchange) by $SO_2$, NO, and $NO_2$ are higher than that for $H_2O$ molecules, thus preventing these strongly bound species from occupying the metal centers with pre-adsorbed CO₂. As a result, SO₂, NO and decomposed NO₂ species are only adsorbed into the *unoccupied* open metal sites and co-exist with CO₂ within MOFs.

### 4.3. DFT modeling of the exchange reaction

In order to calculate the kinetic barrier (activation energy) and study the reaction pathway for the H₂O→CO₂ and SO₂→CO₂ exchange processes, we performed *ab initio* calculations for the CO₂ exchange by H₂O and SO₂ in the Mg-MOF-74.

We start by optimizing the structure and geometry of the Mg-MOF-74 system loaded with CO₂ molecules adsorbed at all the six metal centers. Then, one molecule of H₂O or SO₂ is added to the system separately. It is important to notice that since the MOF is fully loaded with six CO₂ molecules, the added molecules (H₂O or SO₂) are not interacting with the metal centers of the MOF. This situation is expected to recreate the interaction between the added molecules and the adsorbed CO₂ just after the SO₂ and H₂O molecules are introduced into the system. For this case, the system was optimized until the forces acting on each of the atoms were less than 0.001 eV/ Å. Only the configurations with lowest energy were considered, which are shown in Figure 9.

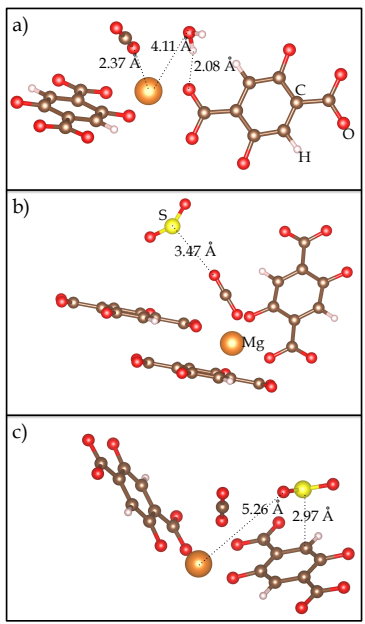

**Figure 9**. Panel (a): adsorption configurations of H₂O on Mg-MOF-74. Panel (b) and (c): adsorption configuration of SO₂ on Mg-MOF-74. In all cases the MOF was fully loaded with CO₂ molecules. Orange, brown, red, yellow and white spheres represent Mg, C, O, S, and H atoms, respectively.

The calculations show that the added H₂O and SO₂ molecules initially bind to secondary adsorption sites, with no direct interaction with the metal centers, as expected. Two secondary adsorption sites were found for SO₂ with binding energies of 43 and 67 kJ/mol, while only one secondary adsorption site was found for H₂O with a binding energy of 49 kJ/mol. The H₂O molecule interacts with the O atoms of the linker establishing a hydrogen bond between the positively charged H atom of the water molecule and the negatively charged O atom of the linker (see Figure 9a). In the case of SO₂ adsorption with a binding energy of 43 kJ/mol, the molecule does not bind to the linker, rather it prefers to stay near the center of the channel above the adsorbed CO₂ molecule (see Figure 9b). On the other hand, for the SO₂ adsorption with a binding energy of 67 kJ/mol, the S atom is responsible for binding the molecule to the carbon atom of the linker (see Figure 9c). Next, we optimized the systems for the cases where the SO₂ and H₂O molecules have replaced one of the CO₂ molecules and attached to the metal center with their oxygen atoms. To study this case, we optimized the MOF structure with five CO₂ molecules and the guest molecule (H₂O or SO₂) adsorbed on each of the metal centers. Then, we added a sixth CO₂ molecule to the system and optimized it until the force acting on each of the atoms were less than 0.001 eV/ Å. Only the lowest energy configurations were considered, see Figure 10. Our results indicate that for the H₂O→CO₂ case, the CO₂ molecule is displaced from the metal center to the secondary adsorption site, located above the linker and close to the water molecule. There, the displaced CO₂ molecule establishes a hydrogen bond with the adsorbed water (calculated binding energy ~ 38 kJ/mol). On the other hand, in the SO₂→CO₂ case, the CO₂ molecule is displaced from the metal center to be adsorbed 3.12 Å above the linker, next to the SO₂ molecule, with a binding energy of 44 kJ/mol. Figure 10 depicts the geometries of the final states.

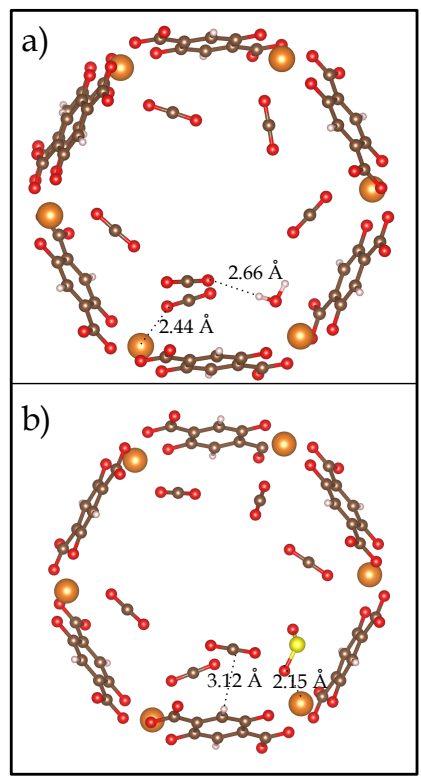

**Figure 10**. Optimized geometries of the fully loaded Mg-MOF-74. Panel (a): H₂O molecule replacing the CO₂ molecule at the metal center. Panel (b): SO₂ molecule replacing the CO₂ molecule at the metal center, with the displaced CO₂ adsorbed on top of the linker. Orange, brown, red, yellow and white spheres represent Mg, C, O, S, and H atoms, respectively.

The configurations shown in Figures 9 and 10 were chosen as the initial and final states for the $H_2O \rightarrow CO_2$ and $SO_2 \rightarrow CO_2$ exchange processes, respectively. It should be mentioned that in the $SO_2 \rightarrow CO_2$ exchange process, the initial state is the one shown in Figure 9b. Based on these initial and final configurations, we then performed NEB (nudged elastic band) calculation[72] to find the kinetic barrier of the transition state; the results are presented in Figure 11. Figure 11a shows that the energy barrier for the $H_2O \rightarrow CO_2$ exchange process has a height of 13 kJ/mol with the transition state shown in the inset. Overall, this is a simple exchange process, where the water molecule only needs to tilt about its H-bonded hydrogen atom to attach to the metal center, overcoming an energy barrier of 13 kJ/mol. It thus lands in the local minimum shown at the 40% mark in Figure 11a. At this point, the $H_2O$ molecule has already displaced the $CO_2$ from the metal center. After this, the $CO_2$ molecule positions itself in the lower energy configuration at the 100% mark of Figure 11a by transitioning trough a series of smaller barriers.

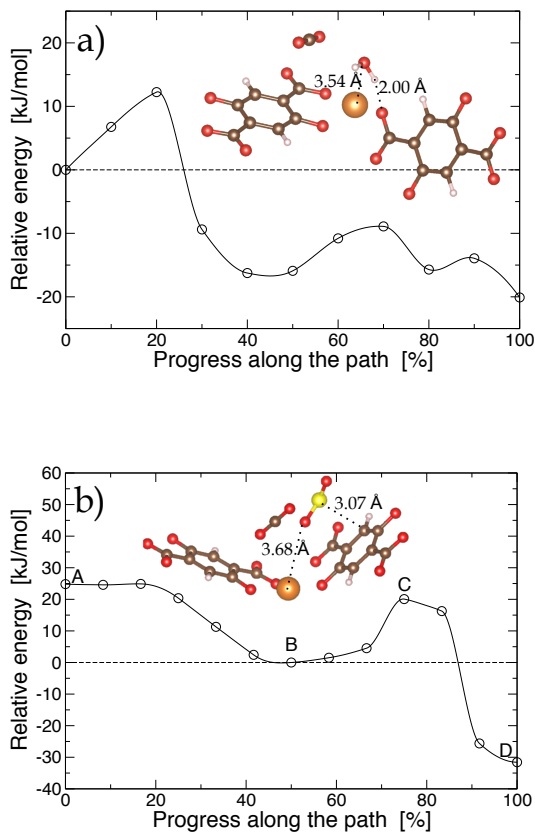

**Figure 11**. NEB calculations. Panel (a): Energy barrier for the $H_2O \rightarrow CO_2$ exchange process. Panel (b): Energy barrier for the $SO_2 \rightarrow CO_2$ exchange process. Capital letters indicate the initial, final, minima, and maximum points. The inset shows the geometry of the transition state at the maximum.

The $SO_2 \rightarrow CO_2$ exchange process has a more complicated path than that of $H_2O \rightarrow CO_2$. This exchange process starts with the geometry configuration depicted in Figure 9b. Then the $SO_2$ molecule overcomes a small energy barrier to arrive in the geometry shown in Figure 9c, i.e. the local minimum at the point marked as **B** in Figure 11b. After this, the $SO_2$ molecule displaces the $CO_2$ molecule from the metal center by overcoming an energy barrier of 20 kJ/mol (the maximum marked as point **C** in Figure 11b, with the transition state shown in the inset), and reaches the global minimum, marked as point **D** in Figure 11b. The transition configuration indicates that $SO_2$ needs to break away from the attractive force of the initial adsorption sites (close to carbon of the benzene ring) and move to push the pre-loaded $CO_2$ away from the metal sites before the global minimum is reached. This minimum corresponds to the geometry shown in Figure 10b. At this point it is important to mention that the kinetic barrier from point **A** to point **B** in Figure 11b is so small (less than 0.5 kJ/mol) that they are negligible for conditions other than cryogenic temperatures. Therefore, for all practical purposes, the $SO_2 \rightarrow CO_2$ exchange process takes places between the geometry configurations depicted in Figure 9c and Figure 10b (points marked as **B** and **D** in Figure 11b). In addition, a vibrational frequency analysis of the transition states (insets in Figure 11) reveals that these geometries have exactly one imaginary frequency, confirming that our NEB calculations have captured the correct transition states.

From the initial and final states depicted in Figures 9 and 10, and the energy paths shown in Figure 11, it is clear that both the $H_2O$ and $SO_2$ molecules gradually position their oxygen atom close to the metal center before overcoming the energy barrier to displace the $CO_2$ molecule away from the metal center. In the $H_2O \rightarrow CO_2$ exchange process, water molecules are initially anchored to the linker's O atoms *nearby* the metal site by a hydrogen bonding interaction. In the $SO_2 \rightarrow CO_2$ exchange process, the $SO_2$ molecule is first adsorbed at the center of the channel. After that, the $SO_2$ molecule overcomes a small energy barrier ($< 0.5$ kJ/mol) and strongly binds to the linker (carbon atom) with its oxygen pointing toward the metal center as shown in Figure 9c. The intermolecular distance between the added molecules ($H_2O$ or $SO_2$) and the metal center is shorter for the $H_2O \rightarrow CO_2$ case than for the $SO_2 \rightarrow CO_2$ case; specifically, the distance between the O atom of the $H_2O$ molecule and the metal center is 4.11 Å (see Figure 9a), whereas that of the O atom of the $SO_2$ molecule is 5.26 Å (see Figure 9c). Finally, after displacing the $CO_2$ molecule from the metal center, both $H_2O$ and $SO_2$ molecules bind strongly to the metal sites via their oxygen atoms. Specifically, the $SO_2$-Mg bond in Figure 10b is 25 kJ/mol stronger than the $H_2O$-Mg bond depicted in Figure 10a.

Based on these results, it is clear that the short distance between the $H_2O$ molecule and the metal center plus the weak hydrogen bonding of the water molecule with the linker (49 kJ/mol) allow the $H_2O$ molecules to easily occupy the metal site and displace the $CO_2$ molecule. On the other hand, the long distance between the $SO_2$ molecule and the metal center plus the strong interaction between the benzene part of the linker and the $SO_2$ molecule (67 kJ/mol) are responsible for the high-energy barrier in the $SO_2 \rightarrow CO_2$ exchange process.

The significantly lower kinetic barrier (13 kJ/mol) for the $H_2O \rightarrow CO_2$ exchange process compared to the higher barrier (20 kJ/mol) for the $SO_2 \rightarrow CO_2$ exchange, explains the observed faster exchange process. Assuming that the $SO_2 \rightarrow CO_2$ and $H_2O \rightarrow CO_2$ exchange processes have the same pre-factor in the Arrhenius equation, the DFT results indicate that the rate constant of the $H_2O \rightarrow CO_2$ exchange process is 17 times greater than that of $SO_2 \rightarrow CO_2$ at room temperature. To estimate the ratio determined from the spectroscopic measurements, the decrease of the $CO_2$ band at 2350 cm$^{-1}$ in Mg-MOF-74 is plotted as a function of time after $H_2O$ and $SO_2$ exposures (see Figure S27). Since the initial step involves moving $CO_2$ molecules adsorbed on the open metal sites at 2350 cm$^{-1}$ to the secondary sites at 2342 cm$^{-1}$, the initial values of the integrated areas of the $CO_2$ bands in both cases are normalized to one. The rate at which the $CO_2$ is leaving the metal centers, which we call $v(H_2O)$ and $v(SO_2)$, is then calculated from the derivative (the "rate") of the two curves. Taking the ratio $(v(H_2O)/v(SO_2))$, we find that in the first 20 seconds, the $CO_2$ decay upon $H_2O$ exposure is 10-25 times faster than the corresponding decay upon $SO_2$ exposure. This is in excellent agreement with the ratio of 17 derived from the DFT calculations.

This theoretical analysis is in good agreement with the experimental results and allows us to conclude that hydrogen bond between the $H_2O$ molecule and nearby oxygen atom of the linker is responsible for anchoring the $H_2O$ molecules close to pre-existing $CO_2$ and displacing the $CO_2$ molecules from the metal centers. To examine the role of hydrogen bonding in other system such as $NH_3 \rightarrow CO_2$, we have optimized the initial co-adsorption state of 6 $CO_2$ + $NH_3$ in Mg-MOF-74, similarly to what was done for the 6 $CO_2$ + $H_2O$ system. The calculations show that the $NH_3$ molecule is again (like $H_2O$) stabilized by a hydrogen bond to the oxygen atom of the linker as shown in Figure 12. It is adsorbed with its H atom located 2.56 Å above the O atom, and the N atom at 4.05 Å from the metal center. Therefore, the $NH_3$ molecules are positioned near the metal centers, similarly to $H_2O$, facilitating (i.e. lowering the barrier for) the displacement of the $CO_2$ molecules out of the metal centers. This is consistent with the observation that ~60% of $CO_2$ at the metal site (2350 cm$^{-1}$) is displaced away from the metal center, towards the secondary site (2342 cm$^{-1}$) within 10 min of $NH_3$ loading (see Figure 8a). This finding for 6 $CO_2$ + $NH_3$ and the results for the 6 $CO_2$ + $H_2O$ system, strongly support the conclusion that hydrogen bonding between the linker and the guest molecules is the key factor that controls the displacement process of a variety of molecules adsorbed at the metal center. Note that $CH_4$ is a non-polar molecule, i.e. its hydrogen is neutral (negligible charge transfer) typical of covalent bonds. Therefore, there is no hydrogen bonding with oxygen of the ligand.[73]

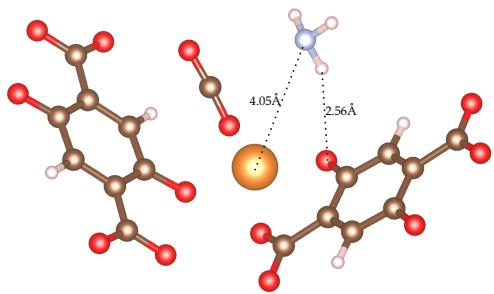

**Figure 12**. Adsorption configuration of the $NH_3$ molecule on Mg-MOF-74 fully loaded with $CO_2$ molecules at the metal centers. Red, brown, grey, blue, and orange spheres denote O, C H, N, and Mg atoms respectively. Parts of the MOF have been removed for visualization purposes.

In summary, the calculations of $H_2O \rightarrow CO_2$ and $SO_2 \rightarrow CO_2$ co-adsorption systems in Mg-MOF-74 show that displacement of one molecule by another within metal organic frameworks is a complex process where the binding energy considerations alone cannot successfully predict the key elements of the reaction. Instead, one must consider the detailed exchange pathway and the activation barrier. The way the attacking guest molecules interact with the host framework is important and determines the ease of the displacement reaction. For example, $SO_2$ (and other molecules without H atoms) cannot be adsorbed at the very nearby oxygen site as water does. It is therefore more energetically costly for $SO_2$ and molecules such as NO or $NO_2$ (i.e., without H atoms) to displace pre-loaded $CO_2$. Since MOFs are complicated systems with diverse chemical environments, the interaction of individual molecules with the pore structure must be fully considered to evaluate the displacement reactions.

**5. Conclusion.**

In this work, the competitive adsorption of $CO_2$ with a series of gases present in flue exhaust ($H_2O$, $NH_3$, $SO_2$, NO, $NO_2$, $N_2$, $O_2$, and $CH_4$) in MOF-74 compounds is examined by combining *in-situ* infrared spectroscopy and *ab initio* calculations. For $CO_2 \rightarrow H_2O$, pre-loaded $CO_2$ is displaced by $H_2O$ through a favorable H-interaction of $H_2O$ with the oxygen atom of the MOF linker. In contrast, other gases like NO and $SO_2$ cannot easily exchange position with $CO_2$, even though their binding energy at the metal sites is higher than that of $CO_2$. These findings suggest that the kinetic energy barrier controls the exchange process, instead of thermodynamic considerations. The DFT results show that molecules such as $H_2O$ are able to establish a hydrogen bond between one of their positive charged H atoms and one of the negative charged O atoms in the linker. On the other hand, molecules such $SO_2$ are not able to do so and remain bound at a distant site. Because the $H_2O$ molecules are initially adsorbed at the nearby secondary adsorption site, next to the metal center, the $H_2O \rightarrow CO_2$ exchange process is facilitated, characterized by a lower energy-barrier path than that of the $SO_2 \rightarrow CO_2$ exchange process. $NH_3 \rightarrow CO_2$ experiments and simulation further support the conclusion that hydrogen bonding is essential to lower the kinetic energy barrier of the displacement process. These findings shed light on the mechanism of competitive co-adsorption, selective adsorption and diffusion. Furthermore, they show that the single indicator of binding energy is not enough to evaluate the performance of the selectivity.

This study clearly highlights that the details of the kinetics (barrier to displace the original molecules) have to be taken into account for co-adsorption and separation processes in MOF materials. The kinetic barriers are affected by many factors such as guest-guest interactions and guest-host interactions. In the case of MOF-74, while the nature of the guest-guest interaction is relevant, the guest-MOF interactions are critical to facilitate the exchange of pre-existing molecules by a guest molecule. Specifically for MOF-74, the kinetic barrier is dominated by the ability of the second molecule to bind to the MOF ligands. This work provides a model for future work to explore different types of MOFs structures and examine the competitive co-adsorption of other relevant gases into MOFs materials.

From practical point of view, the results presented in this work show that water is more detrimental since it replaces the $CO_2$ more efficiently. Even low humidity levels (8 Torr, ~40% RH) are sufficient to remove most of $CO_2$ molecules (>70%) from MOF-74 adsorbents. Therefore, any application of MOF-74 materials for $CO_2$ capture should consider incorporating a water removal step, such as adding a guard bed loaded with silica gel or alumina to the MOFs system.[74] Co-adsorption of $SO_2$, NO, and $NO_2$ is less detrimental since these molecules are not efficient in removing the pre-adsorbed $CO_2$ within MOF-74. Furthermore, their concentration in flue gases is very low, below 1000 ppm. However one must consider the possibility that these molecules could poison the sorption sites and prevent the additional $CO_2$ molecules adsorbing into MOFs.

## ASSOCIATED CONTENT

**Supporting Information**.
Synthesis and Powder X-ray diffraction pattern of MOFs samples; IR spectra of gas phase $CO_2$; Pure $H_2O$ adsorption into Ni, Mg-MOF-74 at low and high humidity; Co-adsorption of $H_2O$ and $CO_2$ adsorption into Ni-MOF-74 at room temperature and Mg-MOF-74 at 50 °C; Co-adsorption of $CO_2$ and $NO_2$, $N_2$, $O_2$, $CH_4$ into MOF-74; Co-adsorption of mixtures of $CO_2$, $SO_2$, $NH_3$ in Mg-MOF-74; Frequency simulation of $H_2O$ and $CO_2$ adsorption into Mg-MOF-74. "This material is available free of charge via the Internet at http://pubs.acs.org."

**Corresponding Author**
Yves J. Chabal: chabal@utdallas.edu

## ACKNOWLEDGMENT
This work was entirely supported by the Department of Energy Grant No. DE-FG02-08ER46491.

**TOC**

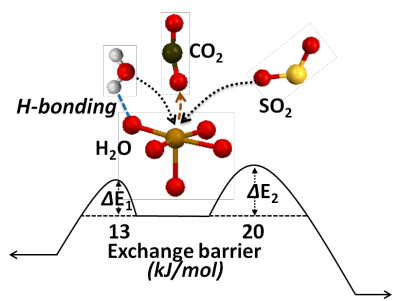